\documentclass[namedreferences]{SolarPhysics}
\usepackage[optionalrh]{spr-sola-addons} % For Solar Physics 

\newcommand{\col}{mono}
\usepackage{graphicx}        % For eps figures, newer & more powerful
\usepackage{amssymb}         % useful mathematical symbols
\usepackage{color}           % For color text: \color command
\usepackage{url}             % For breaking URLs easily through lines
\usepackage{epstopdf}
\usepackage{amsmath}
\newcommand{\half}{{\textstyle\frac{1}{2}}}

\newcommand{\re}{\mathop{\rm Re}\nolimits}

\newcommand{\pderiv}[2]{\frac{\partial#1}{\partial#2}}

\renewcommand{\a}{{\mathbf{a}}}
\renewcommand{\b}{{\mathbf{b}}}

\newcommand{\g}{\mathbf{g}}
\renewcommand{\k}{\mathbf{k}}

\newcommand{\B}{{\mathbf{B}}}

\newcommand{\F}{{\mathbf{F}}}

\newcommand{\U}{\mathbf{U}}

\newcommand{\vdot}{{\mathbf{\cdot}}}
\newcommand{\vcross}{{\mathbf{\times}}}
\newcommand{\grad}{\mbox{\boldmath$\nabla$}}
\newcommand{\bxi}{\mbox{\boldmath$\xi$}}

\newcommand{\thth}{\hspace{1.5pt}}
\newcommand{\curl}{\grad\vcross}
\newcommand{\Curl}{\grad\vcross\thth}

\newcommand\Div{\grad\vdot\thth}

\newcommand{\bv}{Brunt-V\"ais\"al\"a}

\newcommand{\ri}{\mathrm{i}}

  \renewcommand{\le}{\leqslant}

\newcommand{\aap}{\textit{Astron. Astrophys.}}

\newcommand{\apj}{\textit{Astrophys. J.}}

\newcommand{\mnras}{\textit{Mon. Not. Roy. Astron. Soc.}}

\newcommand{\solphys}{\textit{Solar Phys.}}

\begin{document}
%%%%%%%%%%%%%%%%%%%%%%%%%%%%%%%%%%%%%%%%%%%%%%%%%%%%%%%%%%%%%
\begin{opening}

%\title{Helioseismology and the Atmospheric Alfv\'en Wave. I. Direct Numerical Solution}

\title{Three Dimensional MHD Wave Propagation and Conversion to Alfv\'en Waves near the Solar Surface. I. Direct Numerical Solution}

\author{P. S. \surname{Cally}${}^{1,2}$\sep
M. \surname{Goossens}${}^2$}

\institute{${}^1$ Centre for Stellar and Planetary Astrophysics, School of
Mathematical Sciences, Monash University, Victoria 3800, Australia \email{paul.cally@sci.monash.edu.au} (permanent address)\\
${}^2$Centrum voor Plasma-Astrofysica, K. U. Leuven, Celestijnenlaan 200B, 3001 
Heverlee, Belgium \email{marcel.goossens@wis.kuleuven.ac.be}}

\author{M. \surname{Goossens}\email{marcel.goossens@wis.kuleuven.ac.be}}

\runningauthor{P. S. Cally and M. Goossens}

\runningtitle{3D MHD Wave Propagation I}

\begin{abstract}
The efficacy of fast/slow MHD mode conversion in the surface layers of sunspots has been demonstrated over recent years using a number of modelling techniques, including ray theory, perturbation theory, differential eigensystem analysis, and direct numerical simulation. These show that significant energy may be transferred between the fast and slow modes in the neighbourhood of the equipartition layer where the Alfv\'en and sound speeds coincide. However, most of the models so far have been two dimensional. In three dimensions the Alfv\'en wave may couple to the magneto\-acoustic waves with important implications for energy loss from helioseismic modes and for oscillations in the atmosphere above the spot. In this paper, we carry out a numerical ``scattering experiment'', placing an acoustic driver 4 Mm below the solar surface and monitoring the acoustic and Alfv\'enic wave energy flux high in an isothermal atmosphere placed above it. These calculations indeed show that energy conversion to upward travelling Alfv\'en waves can be substantial, in many cases exceeding loss to slow (acoustic) waves. Typically, at penumbral magnetic field strengths, the strongest Alfv\'en fluxes are produced when the field is inclined $30^\circ$--\,$40^\circ$ from the vertical, with the vertical plane of wave propagation offset from the vertical plane containing field lines by some $60^\circ$--\,$80^\circ$.
\end{abstract}

%\begin{acknowledgements}

%\end{acknowledgements}

\end{opening}

%%%%%%%%%%%%%%%%%%%%%%%%%%%%%%%%%%%%%%%%%%%%%%%%%%%%%
\section{Introduction}

\inlinecite{sc06} (hereafter SC) and \inlinecite{cally07} recently presented a generalization of magneto\-hydro\-dynamic (MHD) ray theory that describes mode transmission and conversion between fast and slow magnetoacoustic waves in the context of solar active regions. A uniform, inclined magnetic field was imposed on a plane parallel atmosphere, and magnetoacoustic rays \emph{in the vertical plane containing the field lines} were modelled using the general ray
transmission/conversion formalism of \inlinecite{tkb03}. The ray results were fully confirmed by numerical solution of the wave equations. By remaining in the
plane defined by the magnetic field and gravity though, only the magneto\-acoustic modes are involved. The Alfv\'en wave, the plasma displacement of which is perpendicular to this plane, is entirely decoupled. The purpose of this paper is to extend the wave analysis to the full three dimensions (3D), thereby bringing the Alfv\'en wave into play. This occurs because the Alfv\'en wave displacement vector is rotated out of the horizontal plane, and therefore interacts with the atmosphere's stratification.
The situation differs from the coronal analysis of \inlinecite{melIII} and \inlinecite{melIV}, where field-line curvature and twist are responsible for
tying the magnetoacoustic and Alfv\'en waves together.

Importantly, we integrate from interior to high atmosphere, so as to explore the coupling of helioseismic waves ($p$-modes) with atmospheric oscillations, especially in the vicinity of the Alfv\'en/sound speed equipartition level $z_{eq}$ (typically shallow subsurface in sunspots) where fast/slow conversion occurs. 

Several important lessons may be drawn from our analysis.
\begin{enumerate}
\item The 2D result (SC) that the atmospheric slow wave flux has a pronounced maximum at a particular value of magnetic field inclination from the vertical $\theta$, (typically in the range $20^\circ$--$30^\circ$ depending on spherical harmonic degree $\ell$), survives into 3D, at least out till $|\phi|\lesssim90^\circ$, where $\phi$ is the angle between the vertical plane of wave propagation and the vertical plane in which the field lines lie.
\item There is a distinct maximum in upward Alfv\'enic energy flux in the atmosphere at around $\theta=40^\circ$, $\phi=60^\circ$, with these numbers only weakly dependent on model parameters. The Alfv\'en wave can only be produced by mode conversion in this model.
\item Conversion of magnetoacoustic waves to Alfv\'en waves happens over an extended height range, typically some hundreds of kilometres, as distinct from fast/slow conversion which is normally localized sharply near $z_{eq}$.  
\end{enumerate}

In the interests of brevity, the reader is referred to SC for an extensive discussion of the solar context and significance to local helio\-seism\-ology of MHD waves in strong magnetic fields near the surface, and to several earlier papers exploring MHD mode conversion using a variety of techniques (\opencite{sb92}; \opencite{cb93}; \opencite{cbz}; \opencite{bc97}; \opencite{cb97};\opencite{cally00}; \opencite{cc03}, \citeyear{cc05}).

A 3D analysis from the ray perspective is deferred until Paper II.

%%%%%%%%%%%%%%%%%%%%%%%%%%%%%%%%%%%%%%%%%%%%%%%%%%%%%
\section{Wave Mechanical Formulation}
The linearized wave equations and the necessary boundary conditions are introduced in this section.

\subsection{Equations}
In Section \ref{num_mod}, a broadly realistic solar interior model will be coupled to an overlying isothermal atmosphere. A uniform, inclined magnetic field
\begin{equation}
\B_0 = B_0 \left(\sin\theta\cos\phi,\,\sin\theta\sin\phi,\,\cos\theta\right),
\end{equation}
will be assumed throughout, where $\theta<90^\circ$ is the field inclination from the vertical, and $\phi$ the angle by which it is rotated from the $x$-$z$ plane. The isothermal top adequately represents the chromosphere for our purposes, and conveniently allows us to take advantage of known exact series solutions when selecting top boundary conditions (see Section \ref{AlfBC} and Appendix \ref{frob}). A horizontal and time dependence $\exp[\ri(kx-\omega t)]$ is assumed, with $z$ dependence to be determined, corresponding to wave propagation in the $x$-$z$ plane.

The linearized, adiabatic MHD equations for this scenario may be expressed in terms of the components of the displacement vector $\bxi=(\xi,\,\eta,\,\zeta)$, the sound speed $c$, Alfv\'en speed $a$, gravitational acceleration $g$, and density scale height $H$:
\begin{multline}
a^2\Bigl[\cos \phi  \sin ^2\theta  \sin \phi \, \eta\, k^2-\left(\cos ^2\theta +\sin ^2\theta  \sin ^2\phi \right) \xi \, k^2\\
+\sin \theta  \left(\ri\, k \sin
   \phi  \left(\sin \theta  \sin \phi \,\zeta '-\cos \theta \, \eta '\right)-\cos \theta \cos \phi \, \zeta ''\right)+
  \cos ^2\theta \, \xi ''\Bigr]\\
   +k \left(a^2 k \cos \theta \cos \phi  \sin \theta -\ri\, g\right) \zeta +\left(\omega^2 -c^2k^2\right) \xi +\ri\,c^2 k \,\zeta '=0,   \label{EQ1}
\end{multline}
\begin{multline}
-\left(\ri\, k \xi +\zeta '\right) \frac{c^2}{H} -\ri\, k \cos \phi  \sin ^2\theta  \sin \phi\,  \eta ' a^2+\ri\, k \sin ^2\theta  \sin ^2\phi \, \xi ' a^2+(a^2\sin
   ^2\theta+c^2)  \zeta '' \\
   -\cos \theta  \sin \theta  \sin \phi\,  \eta '' a^2-\cos \theta  \cos \phi  \sin \theta \, \xi '' a^2+\left(\omega ^2-a^2
   k^2 \cos ^2\phi  \sin ^2\theta \right) \zeta \\
   +k \,\xi  \left[k \cos \theta \cos \phi \sin \theta  a^2+\ri
   \left(g+(c^2)'\right)\right]+(c^2)' \zeta '+\ri\, c^2 k\, \xi '=0,   \label{EQ2}
\end{multline}
and
\begin{multline}
\ri\, k \cos \theta \, \xi ' c^2+\cos \theta \, \zeta '' c^2+\left(\omega ^2 \cos \theta -\ri\, g k \cos \phi  \sin \theta \right) \zeta +\omega ^2 \sin \theta
   \sin \phi \, \eta \\
   +\xi  \left(-\frac{\ri\, k \cos \theta \, c^2}{H}+\ri \,g k \cos \theta +(\omega ^2-c^2k^2) \cos \phi  \sin
   \theta +\ri\, k \cos \theta\, {c^2}'\right)\\
   +\left(\ri\, k \cos \phi  \sin \theta\, c^2-\frac{\cos \theta \, c^2}{H}+\cos \theta\,  {c^2}'\right)
   \zeta '=0.                                      \label{EQ4}
\end{multline}
Primes indicate derivatives with respect to $z$.

%%%%%%%
\subsection{Fast, Slow, and Alfv\'en Waves and Radiation Boundary Conditions}  \label{AlfBC}

The top boundary conditions for the fast and slow waves in the superposed isothermal atmosphere are straightforward. The asymptotic controlling factors for the displacements as $z\to\infty$ are respectively (see the Appendix)
\begin{equation}
\exp[\pm kz] \quad\mbox{and}\quad
\exp\left[\left(1-2\,\ri\,k\,H\tan\theta\cos\phi\pm i\sqrt{\frac{\omega^2}{\omega_c^2}\sec^2\theta-1}\right)\frac{z}{2H}\right],
\end{equation}
where $\omega_c=c/2H$ is the acoustic cutoff frequency.

The fast wave is clearly evanescent, with $\exp[-kz]$ the appropriate solution.
The term $2ikH\tan\theta\cos\phi$ in the slow wave controlling factor is purely geometric, accounting for the change in $x$ along an inclined field line: $x-z\tan\theta\cos\phi=\mbox{constant}$ along a field line so the $\exp[ikx]$ dependence cancels this term (recall that the slow wave is rigidly channelled along $\B$ in the high-altitude limit). The $\exp[z/2H]$ term is extinguished by the $\rho^{1/2}$ factors in the kinetic energy density $\half |\rho^{1/2}\dot\bxi|^2$, and similarly in the acoustic energy. The square-root term however is more important. If $\omega>\omega_c\cos\theta$, the slow wave can propagate, and the ``+'' sign is the appropriate choice for it to be upgoing at infinity. On the other hand, for $\omega<\omega_c\cos\theta$, \emph{i.e.}, below the ramp-modified acoustic cutoff frequency, the slow wave is evanescent. In that case too, the ``+'' sign is on the correct choice.

The selection of appropriate top boundary condition for Alfv\'en waves is more subtle. The difficulty is best illustrated by reference to the well-known exact solution for the perpendicular (to both $\B$ and $\g$) velocity $\eta$ in terms of Bessel functions in the 2D isothermal case $\phi=0$:
\begin{equation}
\eta = s^{2 \ri \kappa \tan\theta}\left[A \, J_0(2s\sec\theta)+B\,Y_0(2s\sec\theta)\right],
\end{equation}
where $s=\omega H/a=(\omega H/a_0)\exp(-z/2H)$, $\kappa=kH$ is a dimensionless horizontal wavenumber, and $A$ and $B$ are integration constants. 
If the atmosphere is deemed to extend to $z=\infty$ (\emph{i.e.}, $s=0$), the common practice is to set $B=0$, resulting in a standing wave, the $J_0$ solution alone \cite{fp58,an89}. With this choice, the Alfv\'en wave may not take away energy, contrary to expectations. However, although the $Y_0$ Bessel function diverges as $s\to0^+$ ($z\to\infty$), its energy density is bounded, and so there is no need to dispense with it. The correct choice is actually $B=-\ri A$, whence
\begin{equation}
\eta = A\,s^{2 \ri \kappa \tan\theta}\, H_0^{(2)}(2s\sec\theta),
\end{equation}
\cite{scb84}, where $H_0^{(2)}=J_0-\ri\, Y_0$ is the second Hankel function of order zero, which is well known to represent a wave travelling in the negative $s$ (positive $z$) direction. Although the Alfv\'en wave (travelling at speed $a=a_0\exp(z/2H)$) actually reaches $z=\infty$ in a finite time ($2H/a_0$ from $z=0$), it is what it does there that is important: if it is assumed that it reflects totally, a standing wave quickly results (the $J_0$ solution), but if on the other hand it is assumed lost there, the $H_0^{(2)}$ solution develops. Of course, in reality the isothermal atmosphere (representing the chromosphere) does not extend to infinity. Nevertheless, if we imagine that the wave is lost once it reaches a great height above the solar surface, which is the more physical scenario, the Hankel solution is appropriate. Because we are focussed on the conversion process \emph{per se}, we choose not to complicate issues by including a multi-layer (chromosphere/corona) atmosphere and the resulting resonant partial reflection from the transition region \cite{cally83,scb84}.

The full three-dimensional case is more complicated, because it inherently involves coupling between all three MHD modes: Alfv\'en, fast, and slow. Closed form solutions in terms of special functions are not available. Instead, a Frobenius series expansion is required. This is carried out in the Appendix, where the construction of the outward and inward travelling wave solutions is explained.

With all of this in place, the three top boundary conditions in the numerical solution of the full 3D sixth-order coupled wave equations consist of a matching onto the three physical solutions in the isothermal atmosphere: the evanescent fast wave, the outgoing or evanescent slow wave, and the outgoing Alfv\'en wave.

The bottom magnetic boundary conditions are applied deep enough in the interior that $a\ll c$, and so both the slow and Alfv\'en waves are rigidly field-guided and highly oscillatory with respect to the density scale height. This suggests the radiation condition
\begin{equation}
\left(\pderiv{\ }{t}-\a\,\vdot\grad\right) \Curl\bxi = \mathbf{0},
\end{equation}
where $\a=a\,\widehat\B_0$ is the Alfv\'en velocity. The term in brackets is the field-directed upgoing wave operator, and it is applied to $\Curl\bxi$ so as to suppress the acoustic wave (which is irrotational to a high degree of accuracy) and accentuate the highly oscillatory magnetic waves.

%%%%%%%

\begin{figure}
\centerline{\includegraphics[width=.48\textwidth]{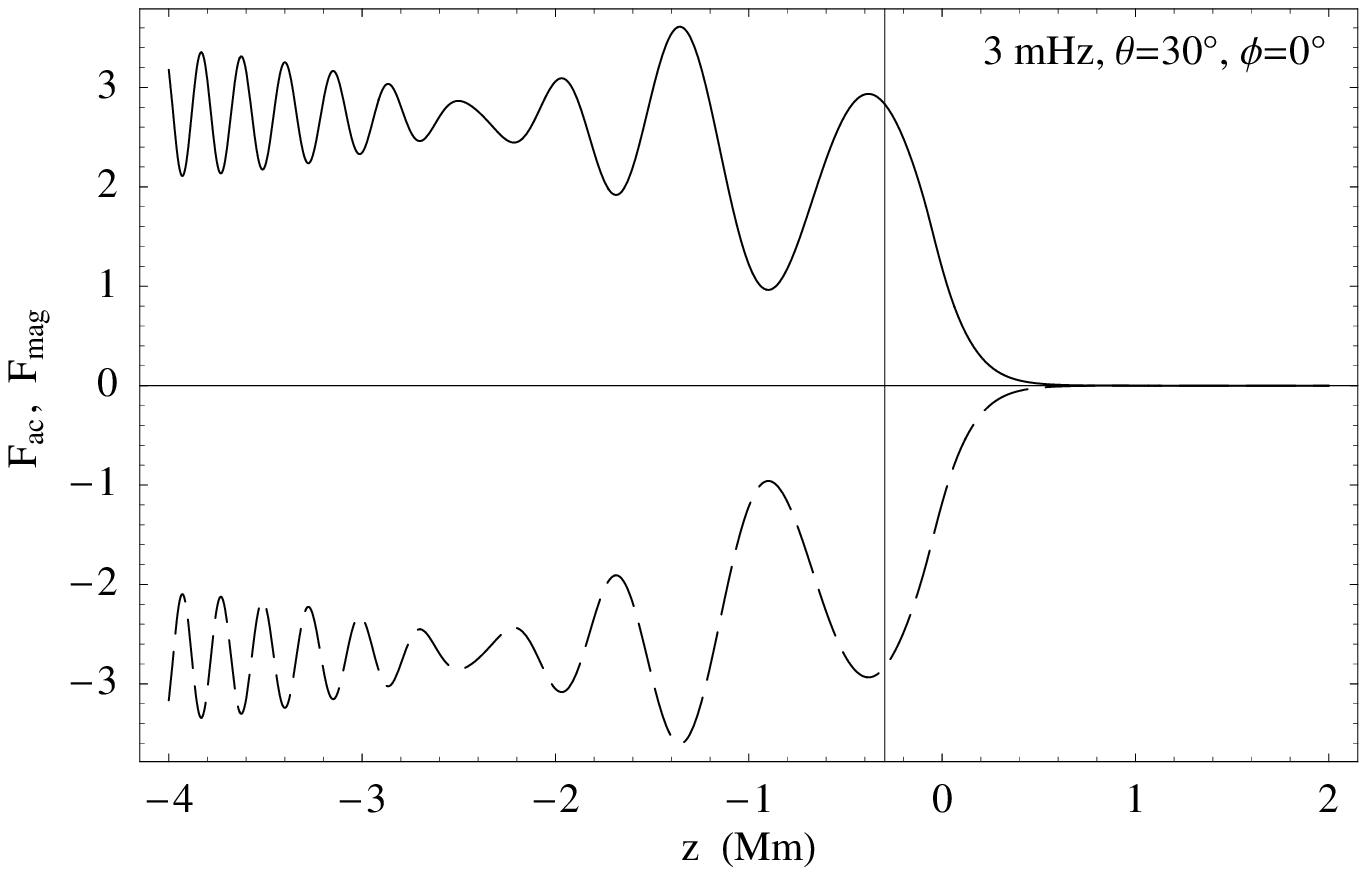}\hfill
\includegraphics[width=.48\textwidth]{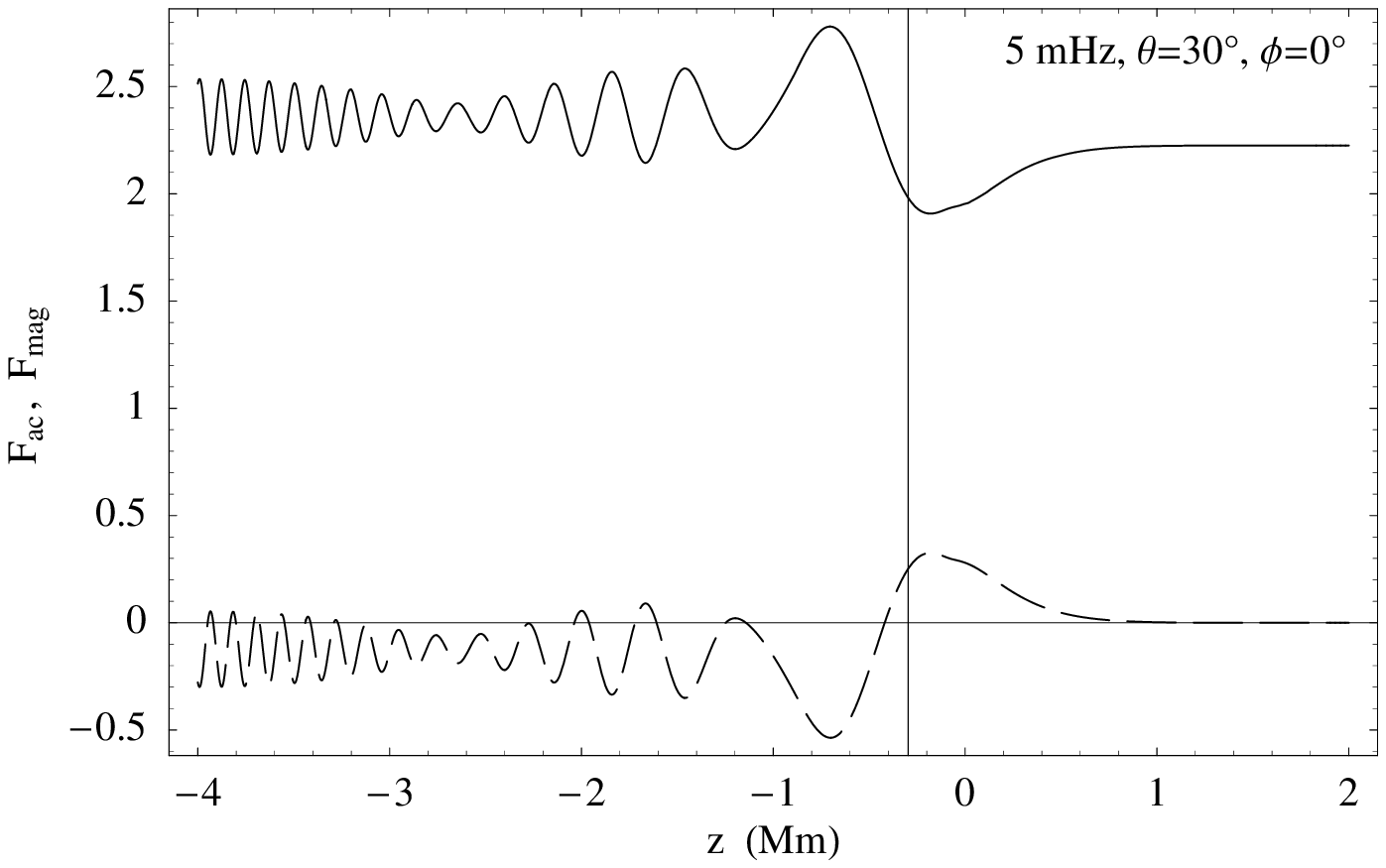}}\smallskip
\centerline{\includegraphics[width=.48\textwidth]{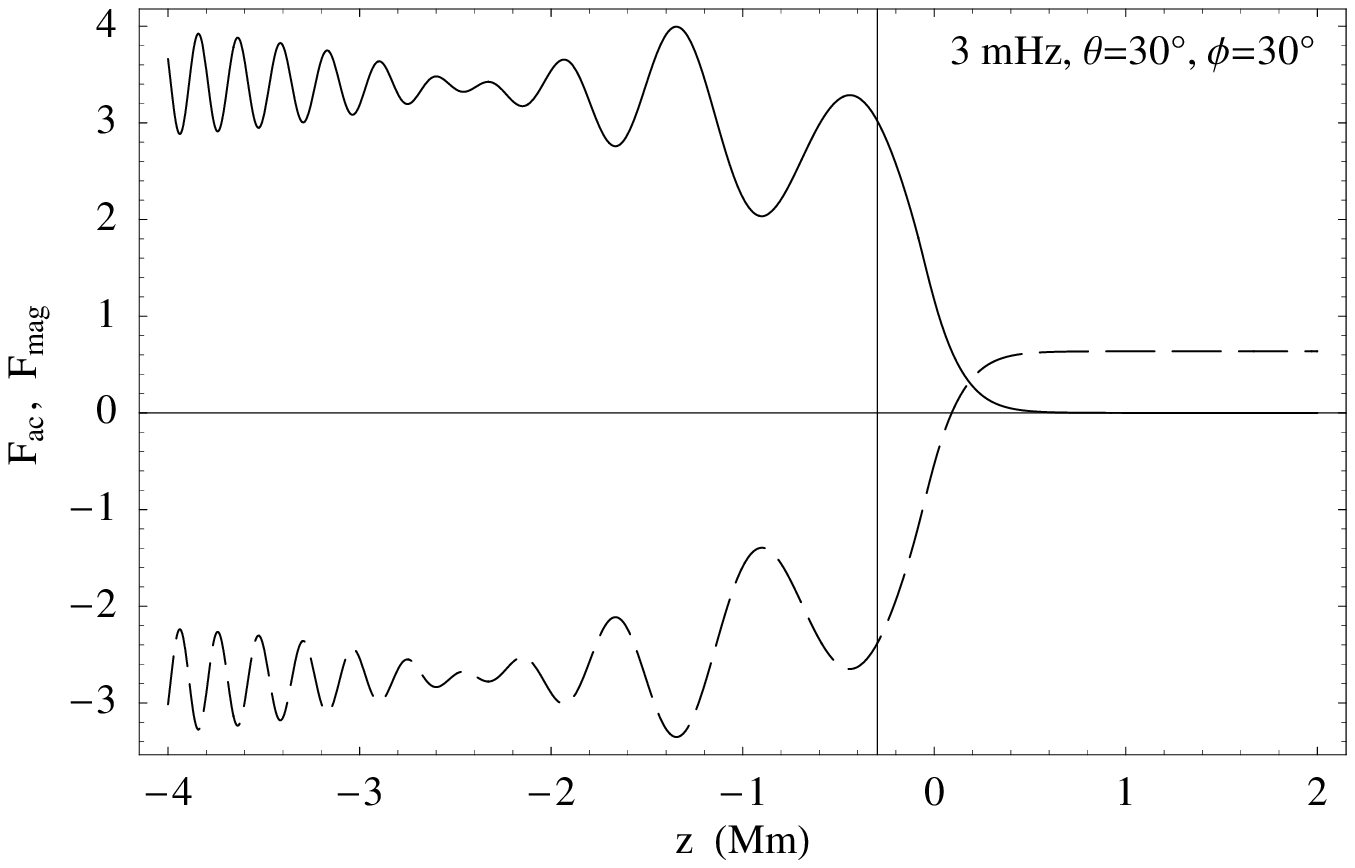}\hfill
\includegraphics[width=.48\textwidth]{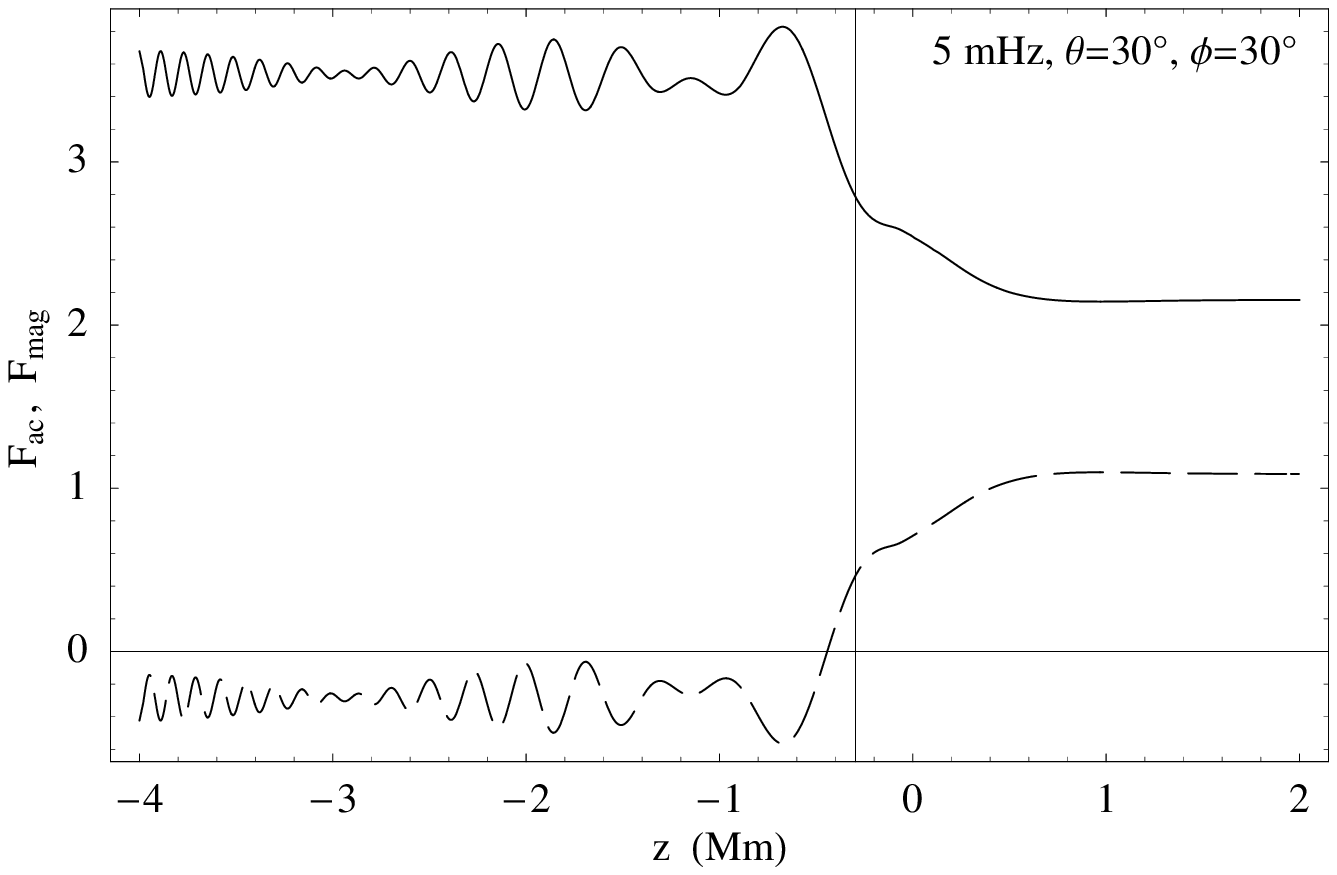}}\smallskip
\centerline{\includegraphics[width=.48\textwidth]{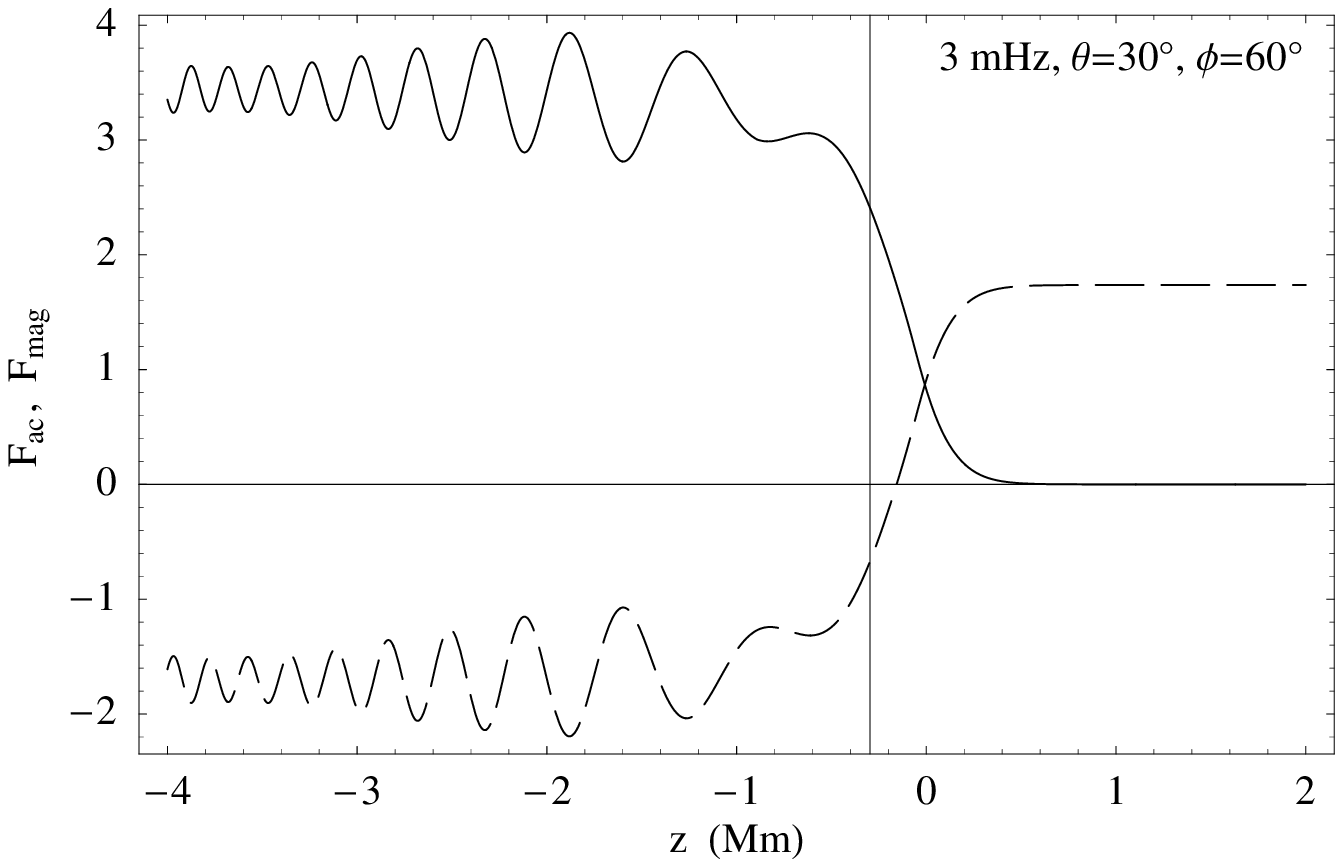}\hfill
\includegraphics[width=.48\textwidth]{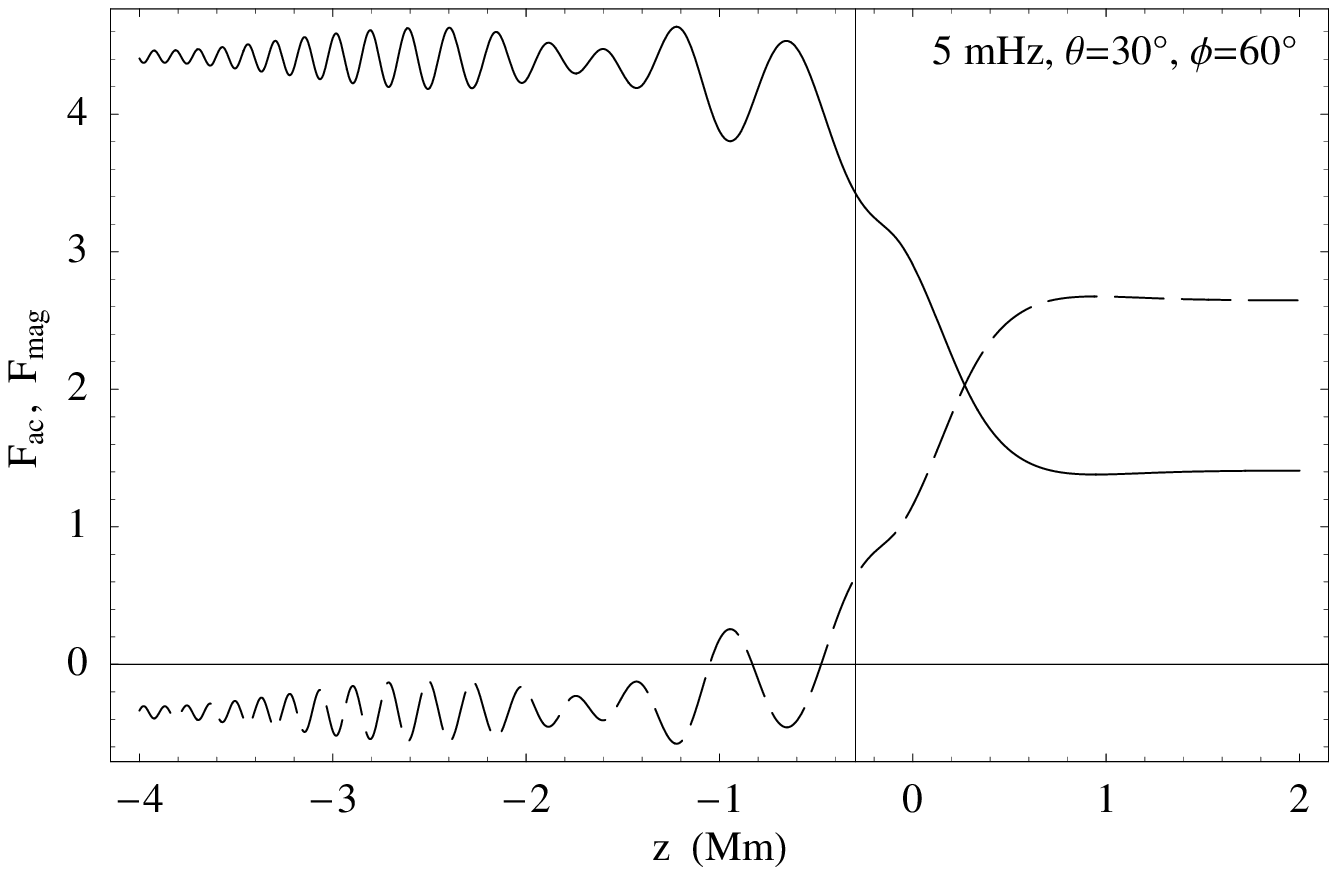}}
\caption{Acoustic (full curve) and magnetic (dashed) wave energy fluxes as functions of height ($z$) over the full computational domain, for two different frequencies: 3 mHz (left panels) and 5 mHz (right panels). In all cases, the 2 kG magnetic field is inclined at $\theta=30^\circ$ to the vertical. The top panels refer to the 2D case ($\phi=0^\circ$), the middle panels to $\phi=30^\circ$, and the bottom panels to $\phi=60^\circ$. The acoustic cavity base is at $z_1=-5$ Mm throughout, corresponding to $k=1.37$ $\rm Mm^{-1}$, or $\ell\approx955$. The vertical axis at $z=-0.297$ Mm indicates the position of the $a=c$ equipartition.}
\label{pf}
\end{figure}

%%%%%%%%%%%%%%%%%%%%%%%%%%%%%%%%%%%%%%%%%%

\section{Numerical Model and Solutions}  \label{num_mod}
We perform a ``scattering experiment'' (in the spirit of \opencite{sb92}) by placing a plane acoustic source at $z_b<0$ beneath the solar surface. The Magnetically Modified Model S (MMMS), a variant on the well-known non-magnetic Model S of \inlinecite{S}, is used below $z=0.5$ Mm, and an isothermal slab above. This model, which reduces the gas pressure (and consequently the sound speed) near the surface to compensate for the introduction of a magnetic pressure, is detailed in the appendix of \inlinecite{sc06}. In practice, we use $z_b=-4$ Mm, though this is not important, providing it is well below the acoustic\,-\,Alfv\'enic equipartition level $z_{eq}\approx -297$ km (for 2 kG field) where $a=c$. At $z_b$, an acoustic wave of frequency $\omega$ is generated which has a natural cavity depth $z_1<z_b$. This determines its horizontal wavenumber ($k$) or alternately spherical harmonic degree ($\ell$). Since $a\ll c$ at $z_b$, the slow mode is predominantly magnetic and transverse, much like the Alfv\'en wave. Radiation conditions are applied on both: no slow or Alfv\'en waves enter the computational region from below. Together with the amplitude of the acoustic driver (set by normalizing the total acoustic energy per unit horizontal area in $z_b<z<0$ to unity), these constitute the three lower boundary conditions. The upper radiation\,-\,evanescence conditions are as prescribed in Section \ref{AlfBC}. Subject to these six boundary conditions (one of which is non-homogeneous), Equations (\ref{EQ1}\,--\,\ref{EQ4}) are solved numerically in $z_b<z<z_t$. We are then free to adjust $\omega$, $z_1$, $\theta$, $\phi$, $B_0$, \emph{etc.}~and to observe the vertical components of acoustic and magnetic wave energy fluxes $F_{ac}$ and $F_{mag}$ reaching the computational top $z_t$, typically set at 2 Mm. 

For reference, we note that the vector wave energy flux is
\begin{equation}
\F=\F_{ac}+\F_{mag}=\re[p_1\mathbf{v}^* +
\mathbf{e}\vcross\b^*],
\end{equation}
\cite{bra74}, where $p_1=-\rho\, c^2\Div\bxi+\rho \,g\,\zeta$ is the Eulerian gas pressure perturbation, $\mathbf{v}=-\ri\,\omega\,\bxi$ is the plasma velocity, $\b=\Curl(\bxi\vcross\B_0)$ is the magnetic field perturbation, and $\mathbf{e}=-\mathbf{v}\vcross\B_0$ is the electric field perturbation. The energy density is made up respectively of kinetic, acoustic, gravitational, and magnetic parts,
\begin{equation}
E= \frac{\rho}{2}\,|\mathbf{v}|^2 + \frac{|p_1|^2}{2\rho c^2} + \frac{\rho}{2}\,N^2 |\zeta|^2 + \frac{|\b|^2}{2\mu},
\end{equation}
where $N$ is the {\bv} frequency. $E$ and $\F$ satisfy the expected conservation equation $\partial E/\partial t +\Div\F=0$ by construction.

Figure \ref{pf} displays the vertical fluxes as functions of $z$ for various cases with a 2 kG magnetic field inclined at $30^\circ$ to the vertical. The top-left panel corresponds to a frequency of 3 mHz with $\phi=0^\circ$, so the Alfv\'en wave is entirely decoupled, and cannot be excited by the acoustic driver at $z_b=-4$ Mm. All magnetic flux in this case is therefore associated with the magnetoacoustic waves, predominantly the slow wave below $z_{eq}$, where it is negative, in accord with the lower radiation boundary condition. Since the (magnetically dominated) fast wave in $z\gg z_{eq}$ is evanescent in all cases, as is the (acoustic) slow wave here since $\omega<\omega_c\cos\theta$ (the acoustic cutoff frequency is 5.2 mHz in the isothermal slab), both magnetic and acoustic fluxes vanish quickly as $z$ increases. In the top-right panel, the frequency is increased to 5 mHz, with other parameters unaltered. In this case, the ramp effect is sufficient to reduce $\omega_c\cos\theta$ below the wave frequency, and so the slow (acoustic) wave propagates upward indefinitely carrying an asymptotically constant positive flux. In the absence of the Alfv\'en wave, the magnetic flux once again vanishes with increasing height.

These two cases are repeated in the middle panels, but for $\phi=30^\circ$. Introduction of 3D coupling has brought in the Alfv\'en wave, as is apparent from the constant positive magnetic flux high in the atmosphere. In the 3 mHz case, \emph{all} upward flux is magnetic, since the acoustic wave is evanescent. At 5 mHz, both acoustic and Alfv\'enic flux contribute substantially.

Finally, the bottom panels repeat the simulations with $\phi=60^\circ$, where the Alfv\'en coupling is even stronger.

\begin{figure}
\centerline{\includegraphics[width=.66\textwidth]{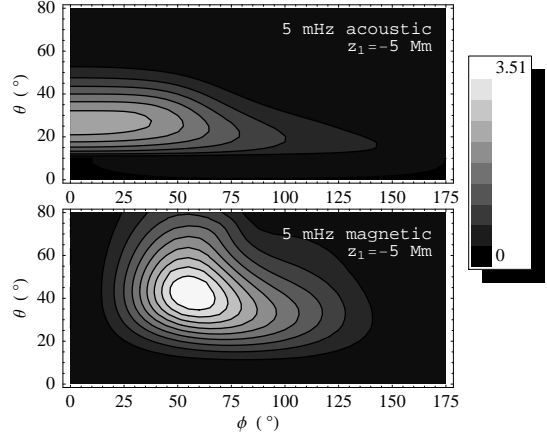}}
\caption{Acoustic (top) and magnetic (bottom) fluxes at $z=z_t=2$ Mm as a function of magnetic field angles $\theta$ and $\phi$ for the 5 mHz, 2 kG case of Figure \ref{pf}. Contours and shading indicate wave energy flux on a linear scale, with white being highest.}
\label{lc5}
\end{figure}

Figure \ref{lc5} illustrates the net outward acoustic and magnetic fluxes for 5 mHz waves in a 2 kG magnetic field of various inclinations ($\theta$) and angles of wave incidence ($\phi$). As shown in SC and \inlinecite{cally07} in the case $\phi=0^\circ$, there is a prominent maximum in acoustic flux at around $\theta=26^\circ$ (for $z_1=-5$ Mm). We now see that this extends out beyond $90^\circ$ in $\phi$, though past say $50^\circ$, the height of the maximum reduces significantly. The surprising breadth of the maximum in the $\phi$ direction is discussed further in Section \ref{conc}. Of course, there is no acoustic flux for $\cos\theta>\omega/\omega_c$ ($\theta\lesssim10.7^\circ$), because of the acoustic cutoff.

\begin{figure}
\centerline{\hspace{-3mm}\includegraphics[width=.57\textwidth]{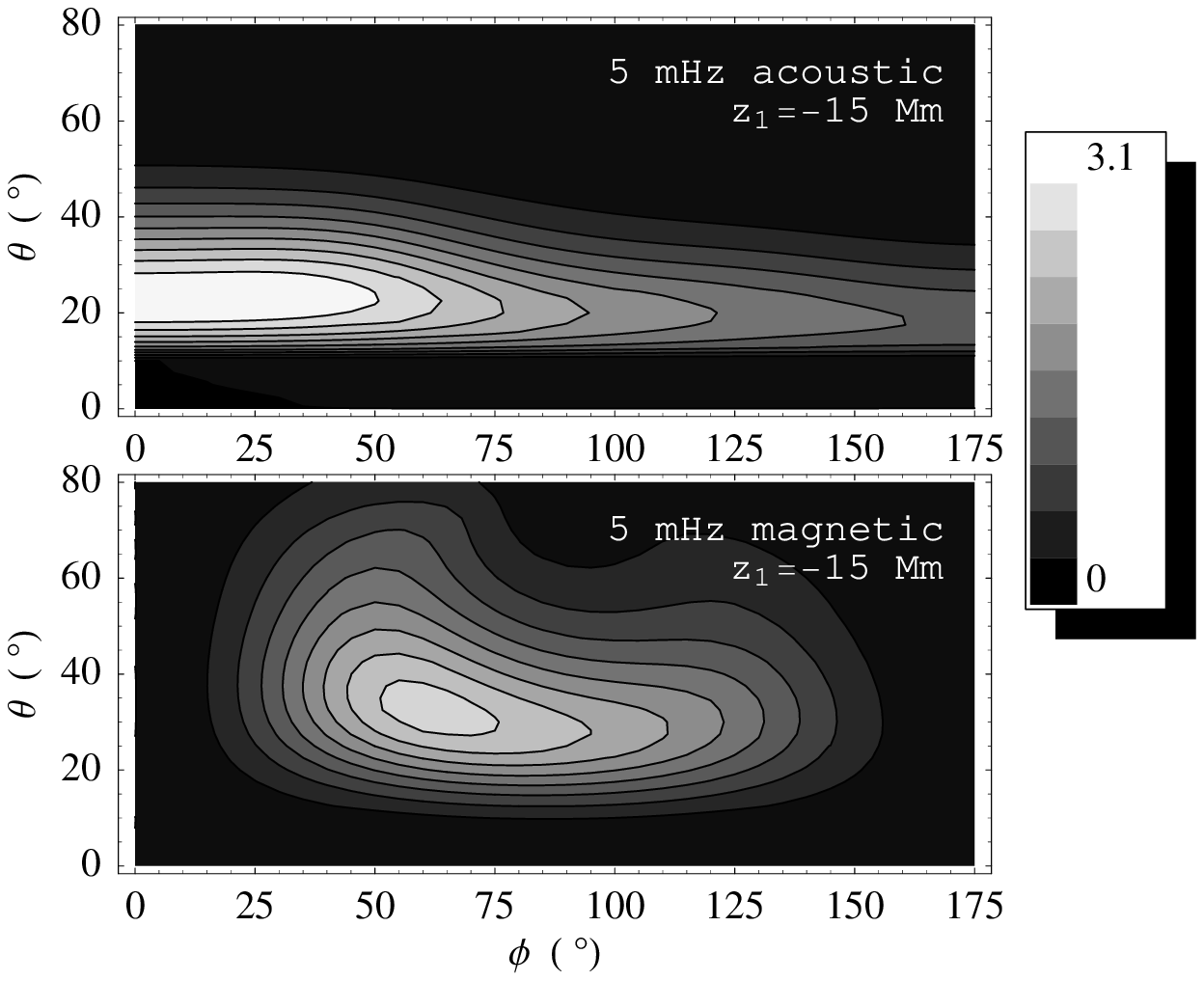}\hspace{-8mm}
\includegraphics[width=.577\textwidth]{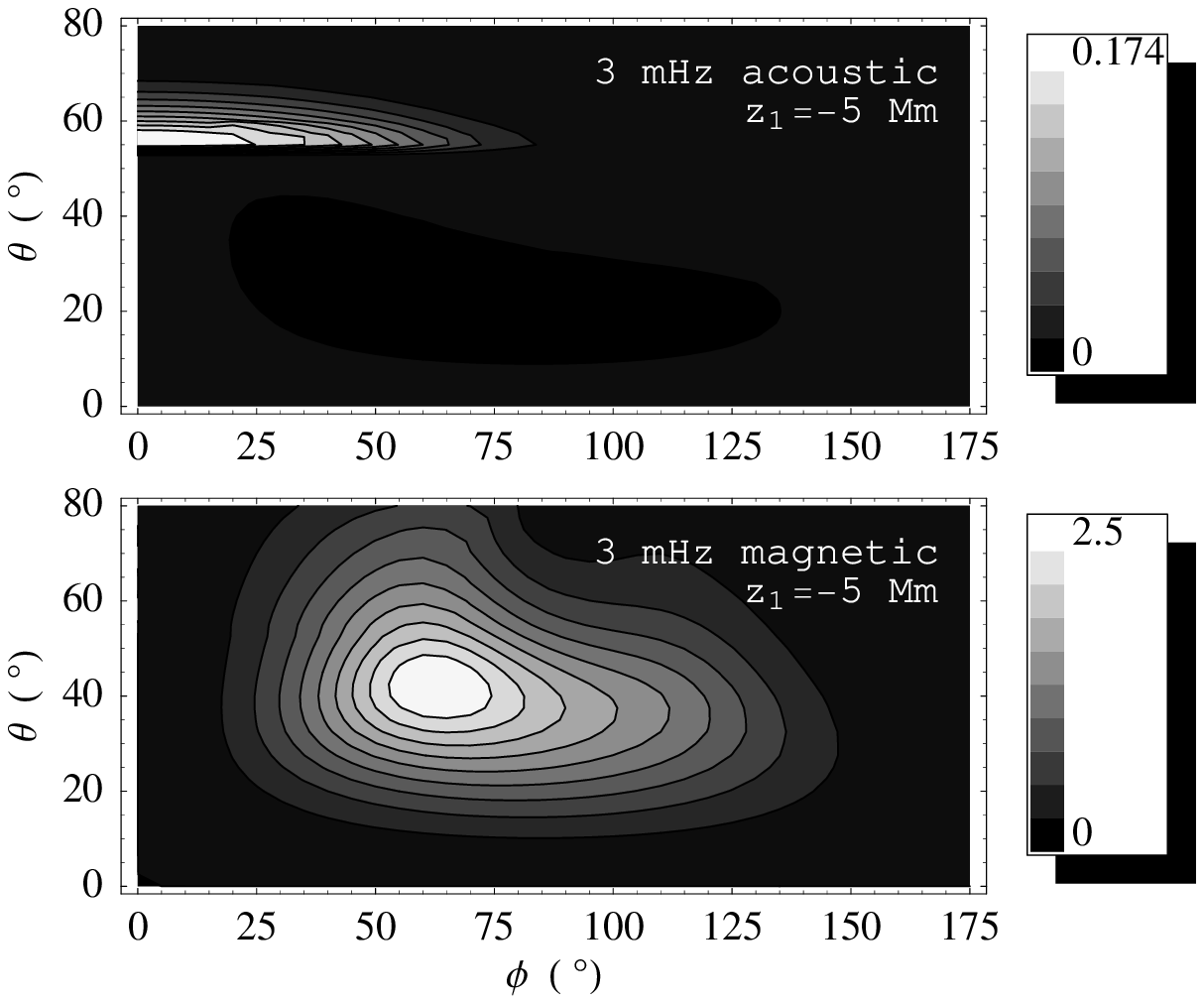}}
\caption{Same as Figure \ref{lc5}, but for different $z_1$ (left) and frequency (right). \emph{Left}: 5 mHz wave with acoustic cavity base $z_1=-15$ Mm, corresponding to $k\approx0.67$ $\rm Mm^{-1}$ ($\ell=468$). \emph{Right}: 3 mHz frequency with $z_1=-5$ Mm ($k=0.79$ $\rm Mm^{-1}$, $\ell=552$). The acoustic and magnetic flux contours are shaded separately in the right panel, since the acoustic flux is much less than the magnetic flux.}
\label{lcOthers}
\end{figure}

On the other hand, Alfv\'enic (magnetic) power peaks at around $\theta=40^\circ$, $\phi=60^\circ$, and has a slightly larger maximum than does the acoustic flux. Very similar results are obtained for 1 kG ($z_{eq}=+53$ km) and 3 kG ($z_{eq}=-613$ km) magnetic field.

%Figure \ref{lc5B3} illustrates the 5 mHz case with $B_0=3$ kG; the acoustic cutoff frequency in the isothermal atmosphere here is 4.94 mHz, \emph{just} low enough for the acoustic wave to propagate at all field inclinations.

%\begin{figure}
%\centerline{\includegraphics[width=.7\textwidth]{lc5B3}}
%\caption{As for Figure \ref{lc5}, but with 3 kG field.}
%\label{lc5B3}
%\end{figure}

Figure \ref{lcOthers} again concerns 2 kG field, but for a deeper acoustic cavity (lower $\ell$), and for a lower frequency, with similar results. In the left panel, it is the acoustic flux that has the slightly higher maximum. In the 3 mHz case (right panel) there is no acoustic flux for $\theta\lesssim53.9^\circ$ because of the acoustic cutoff, and very little thereafter, though the magnetic flux is broadly unchanged.

\begin{figure}
\centerline{\hfill\includegraphics[width=.48\textwidth]{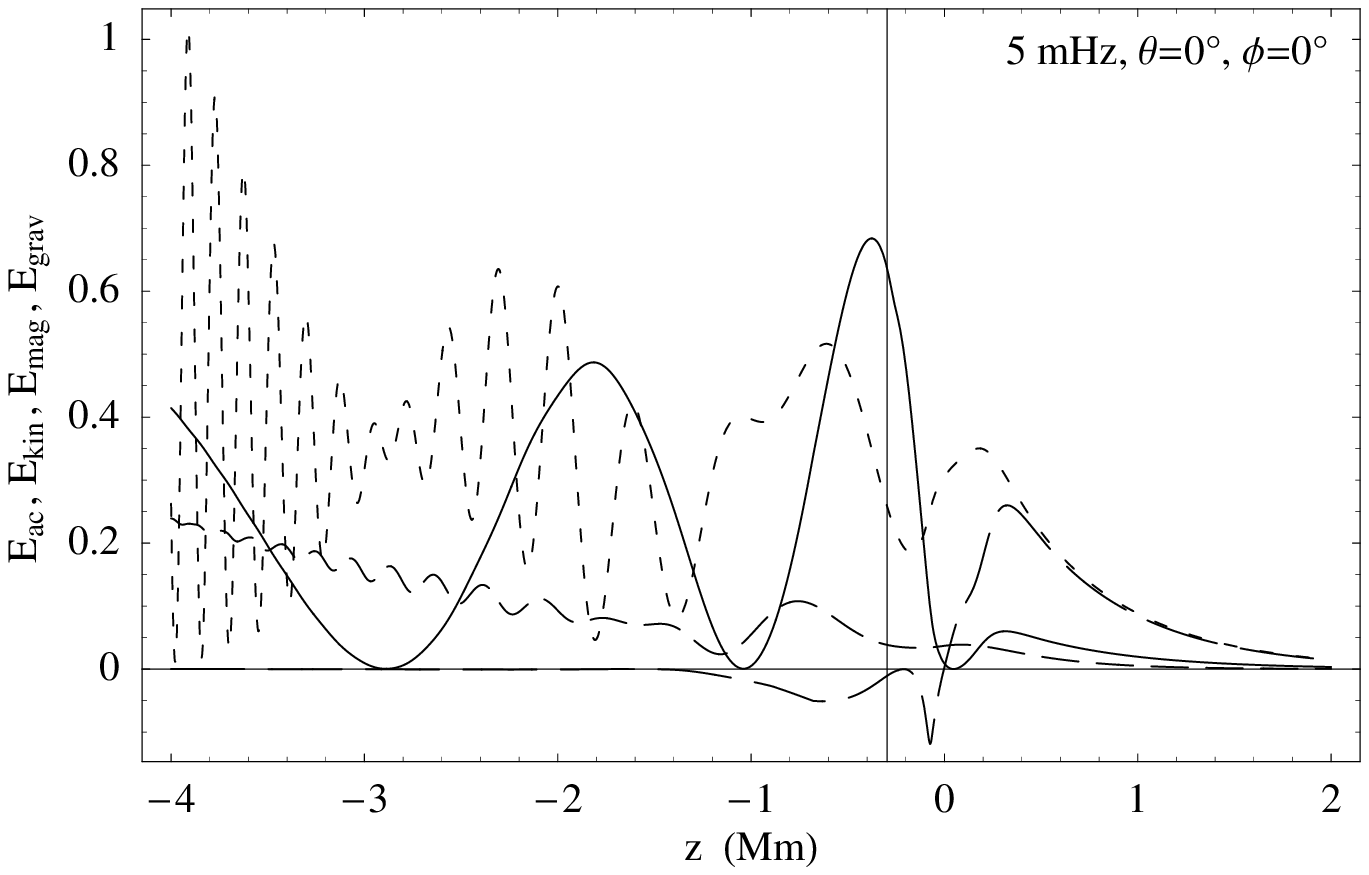}\hfill
\includegraphics[width=.48\textwidth]{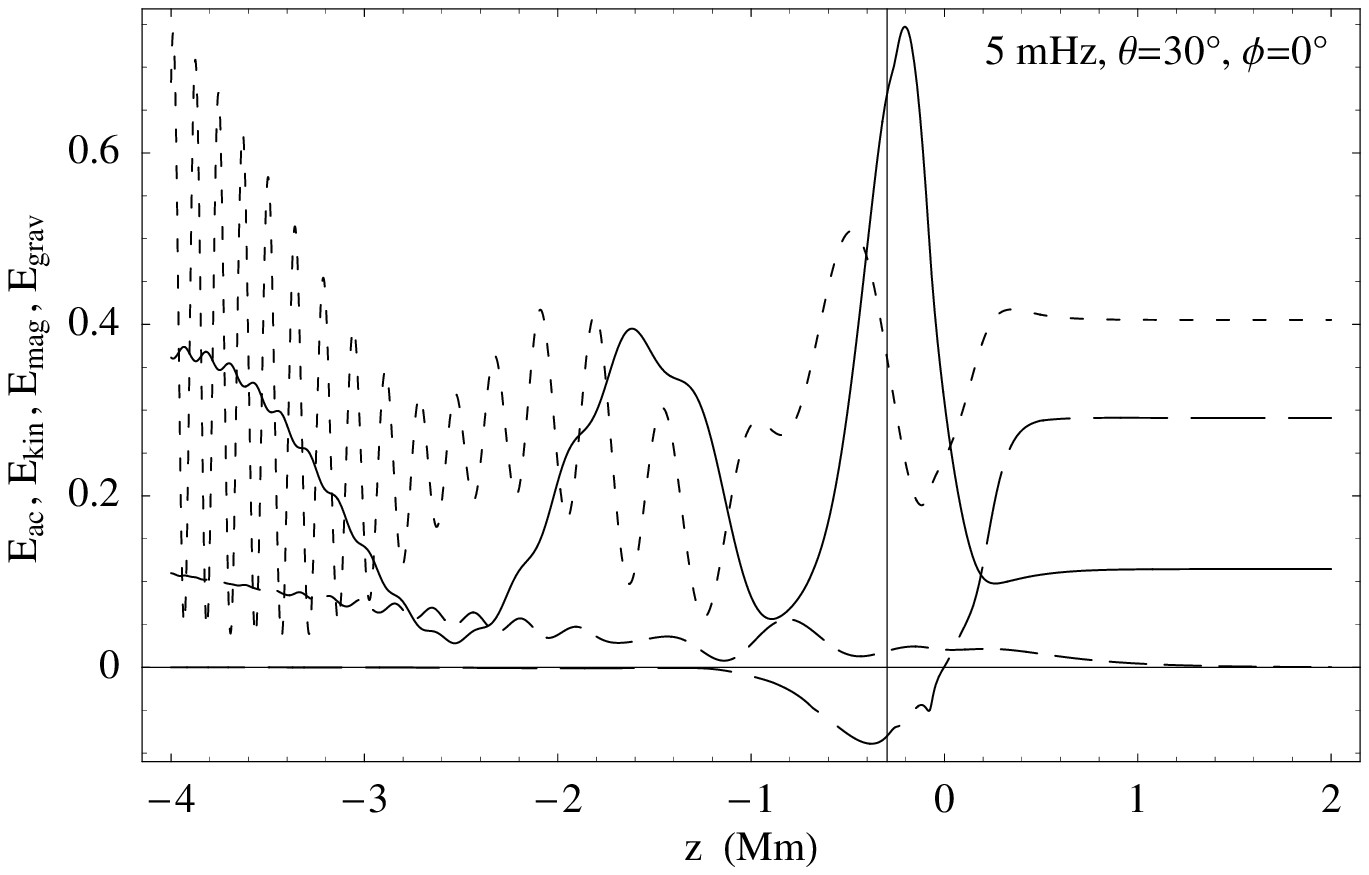}\hfill}\smallskip
\centerline{\hfill\includegraphics[width=.48\textwidth]{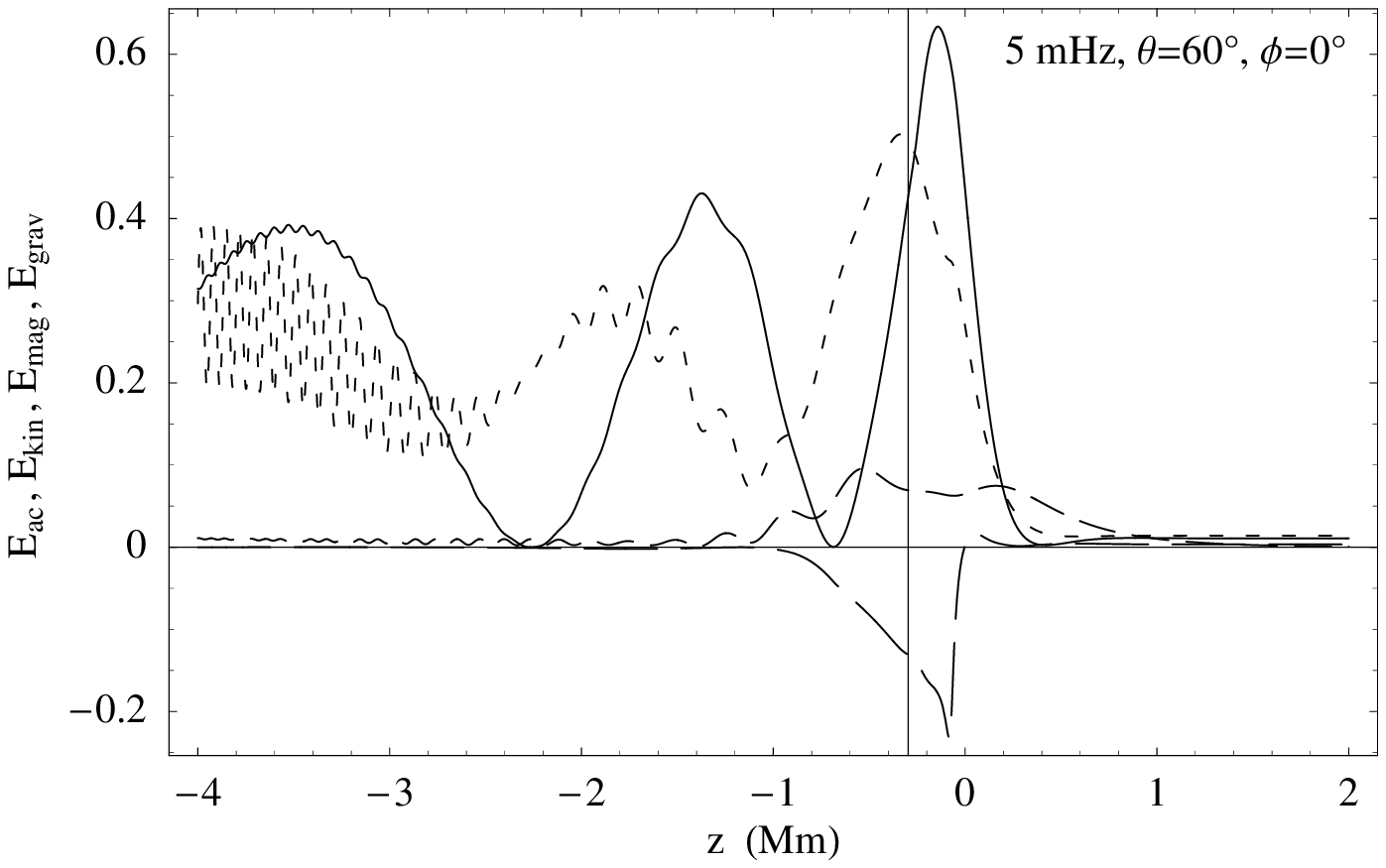}\hfill
\includegraphics[width=.48\textwidth]{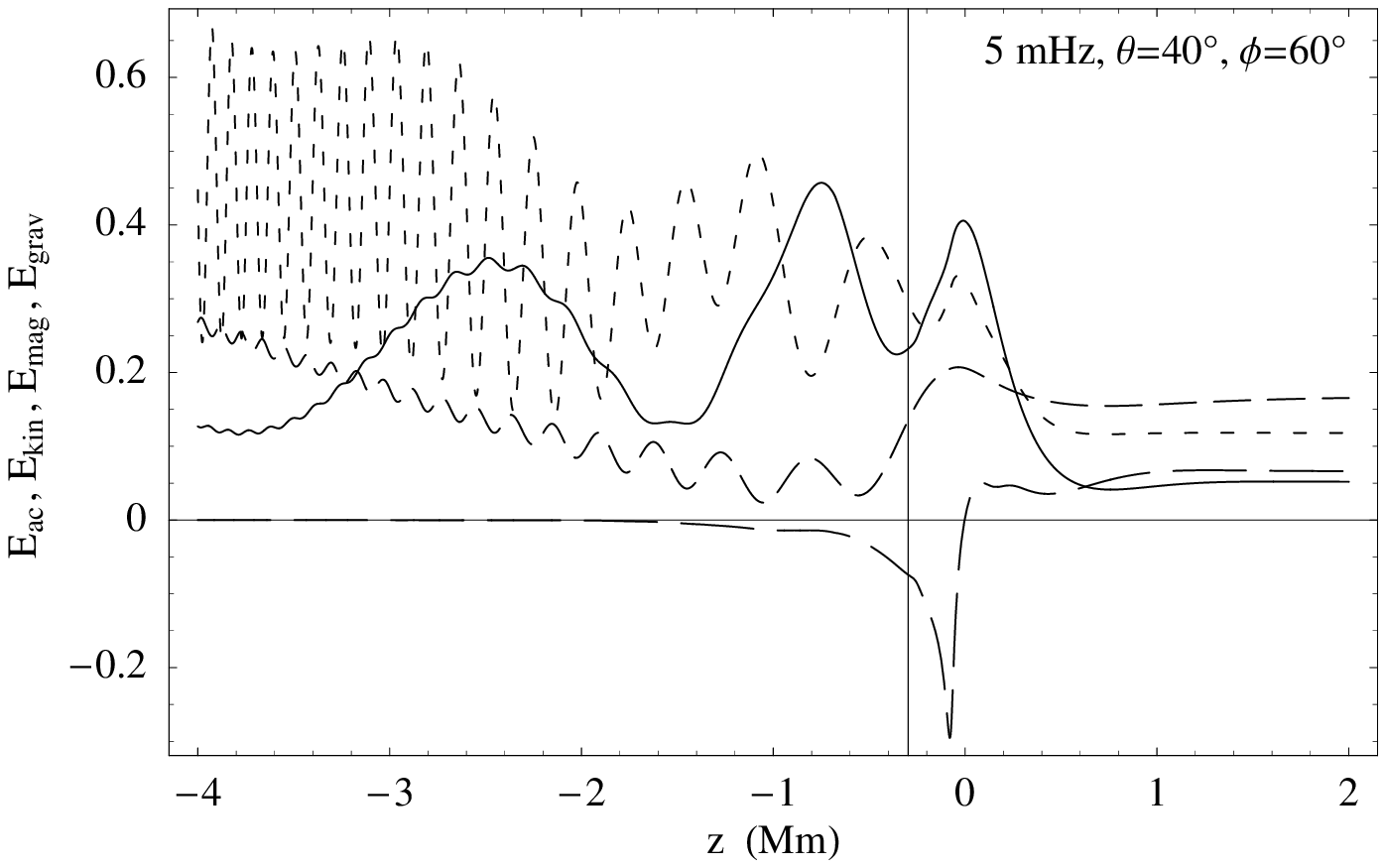}\hfill}\smallskip
\caption{Acoustic (full curve), kinetic (dotted), magnetic (short dashed), and gravitational (long dashed) energy densities for the 5 mHz case of Figure \ref{lc5}, with $\theta$ and $\phi$ as labelled.}
\label{Eall}
\end{figure}

By focussing on the fluxes alone, we have lost sight of the atmospheric fast wave, which is evanescent. However, we may redress this by examining the energy densities. Figure \ref{Eall} displays these for a range of $\theta$ and $\phi$ in the 5 mHz case of Figures \ref{pf} and \ref{lc5}. The first three panels display 2D cases ($\phi=0^\circ$), where the Alfv\'en wave is absent. The magnetic energy density $E_{mag}$ above $z_{eq}$ is therefore a good indication of the strength of the fast wave there. With $\theta=0^\circ$ (both fast and slow waves evanescent), $E_{ac}$ and $E_{mag}$ are of comparable magnitude, though much smaller than $E_{kin}$ and $E_{grav}$. The slow wave is apparently more gravity wave in nature than sound wave; hardly surprising as the {\bv} frequency in the isothermal slab is 4.9 mHz. At $\theta=30^\circ$ the slow wave is travelling, and $E_{ac}\gg E_{mag}$ in the atmosphere. In the highly inclined field case $\theta=60^\circ$ though, where the theory of SC predicts that the coupling to the slow wave is weak, the magnetic energy density dominates between roughly 200 km and 800 km, despite fast wave evanescence. Finally, the 3D case $\theta=40^\circ$, $\phi=60^\circ$, corresponding approximately to the magnetic flux maximum, illustrates how the travelling Alfv\'en wave can dominate the atmospheric energy density. Overall, these figures give a good indication of how much ``leakage'' of $p$-mode energy into the overlying atmosphere we can expect in active regions.

%%%%
\section{Discussion and Conclusions}       \label{conc}

\begin{figure}
\centerline{\includegraphics[width=.6\textwidth]{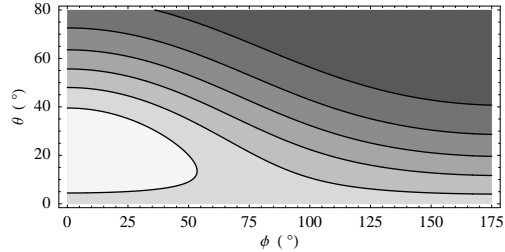}}
\caption{Transmission coefficient $T$ in the $\phi$--$\theta$ plane as given by Equation (\ref{T}) with $|\k|h=1.1$ and $\psi=22^\circ$. The contour levels are 0.9, 0.8, 0.7, \ldots.}
\label{Tcont}
\end{figure}

The greater breadth of the acoustic maxima in the $\phi$ direction compared to the $\theta$ direction exhibited in Figures \ref{lc5} and \ref{lcOthers} may be broadly explained using the concept of \emph{attack angle} discussed in SC, \emph{i.e.}, the angle $\alpha$ between the wavevector $\k$ and the magnetic field $\B_0$. Ignoring the acoustic cutoff and {\bv} frequencies, generalized ray theory predicts an acoustic transmission coefficient of
\begin{equation}
T=\exp\left[-\pi\,|\k|\,h \sec\psi\sin^2\alpha\right]_{a=c},     \label{T}
\end{equation}
where $\psi$ is the angle the wavevector makes to the vertical at the equipartition level $z_{eq}$, and $h=[d(a^2/c^2)/dz]_{a=c}^{-1}$ is the thickness of the $a\approx c$ layer. In 3D,
$\sin^2\alpha=1-(\sin\alpha\cos\phi\sin\psi+\cos\theta\cos\psi)^2$. If we crudely set $|\k|=\omega/c$, then for the case of Figure \ref{lc5}, $|\k|=3.7$ $\rm Mm^{-1}$ and $h=0.29$ Mm, \emph{i.e.}, $kh\approx1.1$. Figure \ref{Tcont} presents contours of $T$ in $\phi$\,--\,$\theta$ space for this case, with $\psi=22^\circ$, and clearly displays a greater breadth in $\phi$ than in $\theta$. If the acoustic cutoff effect is reintroduced, the corresponding flux contours must be squeezed upwards at the bottom to produce a ``blank zone'' below $\theta=10.7^\circ$ (as in Figure \ref{lc5}) and an even more pronounced anisotropy. This is very much in qualitative agreement with the acoustic results of Figures \ref{lc5} and \ref{lcOthers}.

It is clear from the few examples presented in Section \ref{num_mod} that 3D magneto\-acoustic coupling to the Alfv\'en wave is very strong, rivalling and in some cases surpassing the maximum fast-to-slow transmission. Once the angle ($\phi$) between the vertical plane of the wave and the vertical plane of the magnetic field differs, the coupling turns on. It is typically strongest at $\phi=50^\circ$--\,$70^\circ$, and for $\theta=30^\circ$--\,$40^\circ$. This is a significant result, with important implications for our understanding of $p$-mode lifetimes, and atmospheric oscillations above sunspots.

The assumption of uniform magnetic field is made for simplicity and in the spirit of a simple scattering experiment. It is probably quite reasonable for the localized fast\,-\,slow mode conversion. However, the greater breadth of the magneto\-acoustic\,-\,Alfv\'en coupling region suggests that field-line curvature may be relevant to that process. Curvature can certainly provide another Alfv\'en coupling mechanism in its own right.

The upward (Alfv\'en) and downward (slow plus Alfv\'en) magnetic losses illustrated in Figure \ref{pf} indicate a range of relative efficiencies. At fixed $\theta$ and $\phi$ though, it seems that upward loss through the atmosphere is more important at higher frequency. And of course, upward loss is typically of greater relative importance in that $\theta$\,--\,$\phi$ region where $F_{mag}$ is maximal.

The model presented here differs from the earlier \inlinecite{cc05} analysis of the interaction of helioseismic $p$-modes with 3D inclined magnetic field in that (\emph{i}) it consists of a realistic (albeit modified) solar interior model rather than a polytrope; (\emph{ii}) it contains an overlying atmosphere, rather than a vacuum, which allows escape of waves upward as well as downward; and (\emph{iii}) it includes a driver instead of solving for (undriven) eigenmodes. Since our focus is on coupling to atmospheric oscillations, point (\emph{ii}) is especially important. Nevertheless, it is interesting that that study also found substantial (downward propagating) Alfv\'en wave coupling. 

Only the relative magnitudes of the acoustic and magnetic fluxes were quantified in Section \ref{num_mod}. In practical terms though, how big are these fluxes, and how quickly can they deplete the underlying $p$-modes which drive them? Recall that the fluxes were normalized by setting $\mathcal{E}=\int_{z_b}^0 E_{ac}\,dz$ to 1, with $z_b=-4$ Mm in all cases. Now, $F/\mathcal{E}$ has the dimensions of inverse time, and broadly represents the inverse time scale over which the $p$-mode energy in $z_b<z<0$ depletes, assuming it were not being continually replenished by the driver at $z_b$. The maximum fluxes seen in Figures \ref{lc5} and \ref{lcOthers} are of order three in these normalized terms, where time is being measured in kiloseconds. Consequently, the minimum decay time is of order $\tau\sim\frac{1}{3}$ ks $\approx300$ seconds only, \emph{i.e.}, not much more than one period! Of course, this is somewhat of an underestimate of $\tau$, because:
\begin{enumerate}
\item it relies on a particular combination of $\theta$ and $\phi$; only waves travelling in a particular direction and meeting field of a particular inclination will be so strongly affected;
\item we have not integrated the acoustic energy over the entire acoustic cavity $z_1<z\lesssim0$, so the available energy store is larger than indicated by $\mathcal{E}$;
\item there is also a substantial store of \emph{kinetic} energy in the acoustic cavity on which to draw.
\end{enumerate}
Nevertheless, the surface magnetic field and its attendant mode conversion (to slow or Alfv\'en waves) is clearly significant for $p$-modes in active regions.

Figure \ref{pf} helps us understand the nature of the mode conversion. In all cases, we see a broad conversion region starting a little below $z_{eq}$, but extending well above it. This is quite different from the fast\,-\,slow transmission\,-\,conversion process, which is strongly localized around the equipartition level. The reason for this may be discerned using a WKB description, which will be addressed in the forthcoming Paper II.

%%%%%%%%%%%%%%%%%%%%%%%%%%%%%%%%%%%%%%%%%%%%%%%%%%%%%
\begin{acknowledgements}
PSC wishes to acknowledge the generous support of the research council of the K. U. Leuven through the award of a visiting senior postdoctoral fellowship (F/05/088), and the hospitality of the Centre for Plasma Astrophysics, where this work was begun. He also wishes to express his gratitude to Charlie Lindsey, Ashley Crouch, and Ineke De Moortel for very useful discussions during and after the week of the SOHO 19/GONG 2007 meeting in Melbourne.
\end{acknowledgements}

%%%%%%%%%%%%%%%%%%%%%%%%%%%%%%%%%%%%%%%%%%%%%%%%%%%%%%%%%%%%%%%%%%%%%%%%%%%

%%%%%%%%%%%%%%%%%%%%%%%%%%%%%%%%%%%%%%%%%%%%%%%%%%%%%
\appendix

\section{Frobenius Solutions in an Isothermal Atmosphere}   \label{frob}
Exact solutions for the MHD wave equations in an isothermal atmosphere with uniform magnetic field were developed by \inlinecite{zd84}. In the case where the wave vector lies in the vertical plane containing the field, the magnetoacoustic waves decouple from the Alfv\'en wave, and were expressed in terms of Meijer G-functions (although \opencite{cally01}, showed later that the simpler ${}_2F_3$ hypergeometric functions could be used instead). However, in the full 3D case, solution in terms of special functions does not appear to be possible, and Frobenius series were developed instead.

When the two magnetoacoustic waves are coupled to the Alfv\'en waves, the governing differential equations are of sixth order, but there are only five distinct roots of the indicial equation. The Alfv\'enic solutions display a double root, and hence one logarithmic solution. We need this logarithmic solution in order to represent the outgoing Alfv\'en wave. Unfortunately, \inlinecite{zd84} did not pursue the logarithmic case. We therefore develop the necessary solution here.

We adopt the same dimensionless variables as in the case of the decoupled Alfv\'en wave, $\nu=\omega H/c$, $s=\omega H/a=(\omega H/a_0)\exp(-z/2H)$, and $\kappa=k\,H$. The MHD wave equations can be represented as a set of three coupled second-order ODEs in the displacements $\xi(s)$, $\eta(s)$, and $\zeta(s)$, derived from Equations (\ref{EQ1}--\ref{EQ4}). However, computationally, we find it more convenient to write them as a single sixth-order matrix equation\footnote{A partial theory of Frobenius expansion for matrix equations is set out in \inlinecite{rub69}.}
\begin{equation}
s \mathbf{U}' = \mathbf{A} \mathbf{U},   \label{frobeqn}
\end{equation}
where $\mathbf{U}(s)=(\xi,\eta,\zeta,s\xi',s\eta',s\zeta')^T=(\bxi,s\,\bxi')^T$ and
\begin{equation}
\mathbf{A}=
\begin{pmatrix}
\mathbf{0} & \mathbf{I}\\
\mathbf{P} & \mathbf{Q}
\end{pmatrix},
\end{equation}
and where each of the four constituent blocks is $3\times3$. Specifically, 
\begin{equation}
\mathbf{Q}=
\begin{pmatrix}
 2 \ri \kappa  \cos \phi  \tan \theta  & -2 \ri \kappa  \sin \phi  \tan \theta  & 2 \ri \kappa  \left(\frac{s^2 \sec ^2\theta }{\nu ^2}+\tan ^2\theta \right)-2
   \cos \phi  \tan \theta  \\
 0 & 4 \ri \kappa  \cos \phi  \tan \theta  & -2 \sin \phi \tan \theta  \\
 2 \ri \kappa  & 0 & 2 \ri \kappa  \cos \phi  \tan \theta -2
\end{pmatrix},
\end{equation}
and the components of $\mathbf{P}$ are
\begin{align}
P_{11} &=-2 \left(\cos 2 \phi  \tan ^2\theta -1\right) \nu ^2+\left(\left(\frac{4 \kappa ^2}{\nu ^2}-4\right) s^2+4 \kappa ^2-2 \nu ^2\right) \sec ^2\theta \nonumber\\[-8pt]
&\qquad {}+4 \ri(1-\gamma^{-1}) \kappa  \cos \phi  \tan \theta \\
P_{12}&=-2 (\kappa ^2+\nu ^2) \sin 2 \phi  \tan ^2\theta\\
P_{13}&=\frac{4 \ri \left(\cos \phi  \tan \theta  \left(i \gamma  \left(\kappa ^2+\nu ^2\right)+\kappa  \cos \phi  \tan \theta \right) \nu ^2+s^2 \kappa  \sec ^2\theta\right)}{\gamma  \nu ^2}\\
P_{21}&=\frac{4 \ri \sin \phi  \tan \theta  \left(i \gamma  \cos \phi  \tan \theta  \nu ^2+(\gamma -1) \kappa \right)}{\gamma }\\
P_{22}&=-4 \left(s^2 \sec ^2\theta +\left(\nu ^2 \sin ^2\phi -\kappa ^2 \cos ^2\phi \right) \tan ^2\theta \right)\\
P_{23}&=-\frac{4 \sin \phi  \tan \theta  \left(\gamma  \nu ^2-\ri \kappa  \cos \phi  \tan \theta \right)}{\gamma }\\
P_{31}&=\frac{4 \ri (\gamma -1) \kappa }{\gamma }+4 (\kappa^2 -\nu^2 ) \cos \phi  \tan \theta \\
P_{32}&=-4 \nu ^2 \sin \phi  \tan \theta \\
P_{33}&=\frac{4 \ri \kappa  \cos \phi  \tan \theta }{\gamma }-4 \nu ^2,
\end{align}
where $\gamma$ is the ratio of specific heats. Note that $\mathbf{A}(s)=\mathbf{A}_0+\mathbf{A}_2s^2$.

Adopting the standard Frobenius expansion about the regular singular point $s=0$ ($z=\infty$),
\begin{equation}
\mathbf{U}(s) = \sum_{n=0}^\infty \mathbf{u}_n s^{n+\mu},   \label{Ufrob}
\end{equation}
the lowest order balance yields the indicial equation
\begin{equation}
\mathbf{A}_0\mathbf{u}_0=\mu\mathbf{u}_0,
\end{equation}
indicating that $\mu$ is an eigenvalue of $\mathbf{A}_0$ and $\mathbf{u}_0$ is the corresponding eigenvector. The spectrum is
\begin{multline}
\mu\in
\Bigl\{-2 \kappa ,\,2 \kappa ,\,-\ri \sqrt{4 \nu ^2-\cos ^2\theta }\, \sec \theta +2 \ri \kappa  \cos \phi  \tan \theta -1,\\
\ri \sqrt{4 \nu ^2-\cos ^2\theta } \,\sec
   \theta +2 
   \ri \kappa  \cos \phi  \tan \theta -1,\,2 \ri \kappa  \cos \phi  \tan \theta ,\,2 \ri \kappa  \cos \phi  \tan \theta \Bigr\},
\end{multline}
corresponding respectively to the growing (unphysical) fast mode, the evanescent fast mode, the outgoing slow mode (assuming $\nu>\half\cos\theta$), the incoming slow mode, and the Alfv\'en mode. As mentioned before, the Alfv\'en eigenvalue has algebraic multiplicity 2, although its geometric multiplicity (dimensionality of its eigenspace) is only 1.

Beyond the first order, we find that $\mathbf{u}_1=\mathbf{0}$ in all cases, and derive the recurrence relation
\begin{equation}
\mathbf{u}_n = -\left(\mathbf{A}_0-(n+\mu)\mathbf{I}\right)^{-1}\mathbf{A}_2\mathbf{u}_{n-2},
\end{equation}
from which it is apparent that all odd coefficients vanish. This completes the description of the first five solutions.

To discover the sixth solution, we must replace (\ref{Ufrob}) with
\begin{equation}
\mathbf{U}_6(s) = \frac{2}{\pi}\left[\mathbf{U}_5(s)\ln s + \sum_{n=0}^\infty \mathbf{v}_n s^{n+\mu_6}\right].   \label{U6frob}
\end{equation}
The factor $2/\pi$ is just a convenient normalization.
Substituting this into (\ref{frobeqn}) and equating coefficients we find
\begin{gather}
(\mathbf{A}_0-\mu_6\mathbf{I})\mathbf{v}_0=\mathbf{u}_0  \label{v0eqn}\\  
\mathbf{v}_1=\mathbf{0}\\
\left(\mathbf{A}_0-(n+\mu_6)\mathbf{I}\right)\mathbf{v}_n=\mathbf{u}_n-\mathbf{A}_2\mathbf{v}_{n-2},
\end{gather}
where the $\mathbf{u}_n$ are the coefficients in the first Alfv\'enic solution $\mathbf{U}_5$.
Equation (\ref{v0eqn}) yields a generalized eigenvector, in the sense that $\mathbf{A}_0-\mu_6\mathbf{I}$ is singular by definition, with $\mathbf{u}_0$ in its null space, \emph{i.e.}, $(\mathbf{A}_0-\mu_6\mathbf{I})^2\mathbf{v}_0=\mathbf{0}$. We find
\begin{multline}
\mathbf{v}_0=
\Biggl(
\frac{\ri \cos ^3\theta  \sin \theta  \sin \phi }{2 \kappa  \left(1-\sin ^2\theta  \sin ^2\phi \right)},\,\frac{\cos ^2\theta }{2 \gamma  \nu
   ^2},\,\frac{\ri \cos ^2\theta  \cos \phi \sin ^2\theta  \sin \phi }{2 \kappa  \left(\sin ^2\theta  \sin ^2\phi -1\right)},\\
   \frac{\sin ^2\theta  \left((\cos 2 \phi -3) \sin ^2\theta +4\right) \sin 2 \phi }{4 \left(\sin ^2\theta  \sin ^2\phi -1\right)}
  ,\,1-\sin ^2\theta  \sin ^2\phi +\frac{\ri\, \kappa  \cos \theta  \cos
   \phi  \sin \theta }{\gamma  \nu ^2},\\
   \frac{\cos ^3\theta  \sin \theta  \sin \phi }{\sin ^2\theta  \sin ^2\phi -1} \Biggr)^T  .   \qquad\qquad          \label{v0}
\end{multline}
Naturally, this is determined only up to an arbitrary multiple of the Alfv\'enic eigenvector
\begin{multline}
\mathbf{u}_0=\Bigl(-\cos \phi  \sin ^2\theta  \sin \phi ,\,1-\,\sin ^2\theta  \sin ^2\phi ,\,-\cos \theta  \sin \theta  \sin \phi ,\\
-2 \,\ri\, \kappa  \cos ^2\phi  \sin
   ^2\theta  \sin \phi  \tan \theta ,\,
   2 \,\ri \,\kappa  \cos \phi  (1-\sin ^2\theta  \sin ^2\phi ) \tan \theta ,\\
   -2 \,\ri \,\kappa  \cos \phi  \sin
   ^2\theta \sin \phi \Bigr)^T,
\end{multline}
though (\ref{v0}) has been engineered to yield a solution $\U_6$ with zero net energy flux, \emph{i.e.}, it is a standing wave, as we shall see shortly.

The remaining task is to determine which combinations of $\mathbf{U}_5$ and $\mathbf{U}_6$ represent incoming and outgoing Alfv\'en waves at $z=\infty$. To do this, we evaluate the field-aligned component of the Poynting flux, $F_\parallel=\widehat{\mathbf{B}}_0\vdot\mathbf{F}_{mag}$, where $\mathbf{F}_{mag}=\re[i\omega (\bxi\vcross\mathbf{B}_0)\vcross\curl(\bxi^*\vcross\,\mathbf{B}_0)]$, as $s\to0^+$, which only involves the $n=0$ coefficients. Letting $\mathbf{U}=C_5\mathbf{U}_5+\ri\, C_6\mathbf{U}_6$, we find
\begin{equation}
F_\parallel = -\half F_0\cos\theta \left(C_5^*C_6+C_5C_6^*\right)\left(1-\sin^2\theta\sin^2\phi\right),  \label{Fpar}
\end{equation}
where $F_0={B_0^2 c \,\nu/(\pi H^2)}$. Notice that $F_\parallel=0$ if either $C_5=0$ or $C_6=0$, indicating that $\mathbf{U}_5$ and $\mathbf{U}_6$ are both standing waves. Also note that $F_\parallel$ is invariant under the transformation $C_5\to C_5+\ri\,\alpha\, C_6$, $\alpha\in\mathbb{R}$, and similarly for $C_6$.

Setting $C_5=A_++A_-$ and $C_6=A_--A_+$, we may alternatively represent the general Alfv\'en wave as a linear combination of upgoing and downcoming modes, $\U=A_+\U_+ +A_-\U_-$, where 
$\U_\pm=(1\mp \ri\,\alpha)\U_5\mp \ri\,\U_6$,
with total flux
\begin{equation}
F_\parallel = F_0\left[|A_+|^2-|A_-|^2\right]\cos\theta\left(1-\sin^2\theta\sin^2\phi\right).
\end{equation}
The vertical flux is $F_z=F_\parallel\cos\theta$. The real coefficient $\alpha$ does not affect the energy flux, but does contribute to the energy density and the solution matchings. To determine it, we look for guidance in the two dimensional case.

\subsection{2D Case}
Consider the 2D case $\phi=0$, where $\xi=\zeta=0$ and only the transverse displacement $\eta$ remains:
\begin{gather}
\eta_5=s^{2\ri\kappa\tan\theta}J_0(2s \sec\theta),\\[6pt]
\eta_6=s^{2\ri\kappa\tan\theta}\left[Y_0(2s\sec\theta)+\left(
\frac{\cos^2 \theta }{ \pi  \gamma  \nu ^2}-\frac{2\ln \sec \theta }{\pi }-\frac{2
   \mathcal{C}}{\pi }\right)J_0(2s \sec\theta)\right],
\end{gather}
where $\mathcal{C}=0.577216\ldots$ is Euler's constant. Clearly, $J_0(2s\sec\theta)$ and $Y_0(2s\sec\theta)$ are linearly independent solutions, as expected. The corresponding outgoing wave solution then takes the familiar Hankel function form
\begin{align}
\eta_+ &= s^{2\ri\kappa\tan\theta}\left[(1-\ri\beta)J_0(2s \sec\theta) -\ri\,Y_0(2s\sec\theta)\right]\\
& = s^{2\ri\kappa\tan\theta} \left[H_0^{(2)}(2s\sec\theta)-\ri\,\beta\, J_0(2s\sec\theta)\right],
\end{align}
where $\beta=\alpha+
(\frac{\cos^2 \theta }{ \pi  \gamma  \nu ^2}-\frac{2\ln \sec \theta }{\pi }-\frac{2\mathcal{C}}{\pi })$. It is well-known\footnote{This is verified using the large argument asymptotic behaviour \cite{as}, $H_0^{(2)}(x)\sim\sqrt{2/(\pi x)}\exp[-\ri(x-\pi/4)]$.} that the $H_0^{(2)}$ Hankel function represents a ``pure'' wave travelling in the negative $s$ (positive $z$) direction, so it is apparent that we must set $\beta=0$. This determines $\alpha$ in the 2D case.

\subsection{3D Case}In 3D, we do not have the luxury of closed-form solutions from which the large $s$ asymptotic behaviour may be determined. In fact, this would not even be appropriate, as the Alfv\'en wave undergoes coupling to the magnetoacoustic waves around $s=\mathcal{O}(1)$. We are trying to find the real $\alpha$ which delivers a pure outgoing wave \emph{at infinity}, despite it not exhibiting sinusoidal behaviour there. But how do we identify a ``pure'' wave in this regime? To do this, we can again use the Hankel function, or at least its asymptotic behaviour as $s\to0^+$,
\begin{equation}
H_0^{(2)}(2s\sec\theta)\sim 1 - \frac{2\,\ri}{\pi}\,\left(\ln s +\ln\sec\theta+\mathcal{C}\right) +\mathcal{O}(s^2\ln s).         \label{H0}
\end{equation}
(As $s\sec\theta$ is distance along a field line, independent of $\phi$, it is the appropriate spatial coordinate for the field-guided Alfv\'en wave.)\phantom{.} A pure outgoing wave should have this asymptotic structure.

Now, it is easily confirmed that the asymptotic polarization direction of $\bxi_+$ as $s\to0$ is $\mathbf{d}=(-\cos\phi  \sin ^2\theta  \sin \phi ,\,1-\sin ^2\theta  \sin ^2\phi ,\,-\cos \theta  \sin \theta  \sin \phi )$, which is perpendicular to $\B_0$ as expected, and that the small $s$ behaviour of displacement in this direction is
\begin{equation}
\mathbf{d}\,\vdot\,\bxi_+ \sim 1-\ri\,\alpha -\frac{2\,\ri}{\pi}\ln s -\frac{\ri\,\cos^2\theta}{\pi\gamma\nu^2},
\end{equation}
where an arbitrary normalization has been suppressed. Comparing this with Equation (\ref{H0}), we again infer that
\begin{equation}
\alpha = \frac{2\ln \sec \theta }{\pi }+\frac{2\mathcal{C}}{\pi }-
\frac{\cos^2\theta}{\pi\gamma\nu^2},
\end{equation}
as in 2D.

In summary, the three required radiation solutions at large $z$ are $\U_2$, $\U_3$, and $\U_+=(1-\ri\,\alpha)\U_5-\ri\,\U_6$. Physical solutions must match to a linear combination of these. In practice, we apply the boundary conditions at $z=2$ Mm, where $s=\omega H/a\sim\mathcal{O}(10^{-4})$ or smaller, so convergence of the series is very rapid.

%\end{article}
\end{document}